\documentclass[aps,pra,superscriptaddress,twocolumn,showpacs]{revtex4}%
\usepackage{bm}
\usepackage{graphicx}
\usepackage{amsmath}
\usepackage{amsthm}
\usepackage{wasysym}
\usepackage[utf8]{inputenc}
\usepackage{amssymb}
\usepackage{epstopdf}
\usepackage{txfonts}
\usepackage{amsfonts}

\DeclareMathOperator*{\Tr}{Tr} 
 
\DeclareMathOperator{\diag}{diag}
\DeclareRobustCommand\openzero{\leavevmode\hbox{0\kern-.55em0}}
\mathchardef\minus="002D

\newcommand{\ket}[1]{|{#1}\rangle}
\newcommand{\bra}[1]{\langle{#1}|}

\begin{document}
\title{Quantum fidelity for arbitrary Gaussian states}
\author{Leonardo Banchi}
\affiliation{Department of Physics and Astronomy, University
College London, Gower Street, London WC1E 6BT}
\author{Samuel L. Braunstein}
\affiliation{Department of Computer Science, University of York,
York YO10 5GH, United Kingdom} \affiliation{York Centre for
Quantum Technologies (YCQT), University of York, York YO10 5GH,
United Kingdom}
\author{Stefano Pirandola}
\affiliation{Department of Computer Science, University of York,
York YO10 5GH, United Kingdom} \affiliation{York Centre for
Quantum Technologies (YCQT), University of York, York YO10 5GH,
United Kingdom}

\begin{abstract}
We derive a computable analytical formula for the quantum fidelity
between two arbitrary multimode Gaussian states which is simply
expressed in terms of their first- and second-order statistical
moments. We also show how such a formula can be written in terms
of symplectic invariants and used to derive closed forms for a
variety of basic quantities and tools, such as the Bures metric,
the quantum Fisher information and various fidelity-based bounds.
Our result can be used to extend the study of continuous-variable
protocols, such as quantum teleportation and cloning, beyond the
current one-mode or two-mode analyses, and paves the way to solve
general problems in quantum metrology and quantum hypothesis
testing with arbitrary multimode Gaussian resources.
\end{abstract}

\pacs{03.65.-w, 42.50.-p, 03.67.-a, 03.67.Hk} \maketitle





\section{Introduction}

The quantification of the similarity between two quantum states is
a crucial issue in quantum information theory~\cite{Nielsen,Wilde}
and, more generally, in the entire field of quantum
physics~\cite{Sakurai}. Among the various notions, that of quantum
fidelity~\cite{BuresFID,Jozsa,Uhlmann} is perhaps the most
well-known for its use as a quantifier of performance in a variety
of quantum protocols. Quantum fidelity is the standard tool for
assessing the success of quantum teleportation~\cite{Bennett,
Zeilinger, Bra98, CVtelepo, teleREV}, where an unknown state is
destroyed in one location and reconstructed in another (see
Ref.~\cite{telereview}\ for a recent review). In quantum
cloning~\cite{Buzek,lin,Cerf1,braclone,fiu}, where an unknown
state is transformed into two or more (imperfect) clones, quantum
fidelity is the basic tool to quantify the performance of a
quantum cloning machine. Quantum fidelity plays a central role in
quantum metrology~\cite{SamMETRO,Paris}, where the goal is to find
the optimal strategy to estimate a classical parameter encoded in
a quantum state. Similarly, it is important in quantum hypothesis
testing~\cite{QHT,QHT2}, where the aim is to optimize the
discrimination of quantum hypotheses (states or channels).

An important setting for all the above tasks is that of
continuous-variable systems~\cite{WeeRMP,SamRMP}, which are
quantum systems with infinite-dimensional Hilbert spaces, such as
the bosonic modes of the electromagnetic field, described by
position and momentum quadrature operators. For these systems,
Gaussian states~\cite{WeeRMP} are the most typical quantum states
in theoretical studies and experimental implementations, so
quantifying their similarity is of paramount importance. The
derivation of a simple formula for the quantum fidelity between
two arbitrary bosonic Gaussian states is a long-standing open
problem with a number of partial solutions accumulated over the
years. We currently know the solutions for one
mode~\cite{fidG2,fidG3,fidG4} and two
modes~\cite{marian_uhlmann_2012}. A simple formula for multimode
Gaussian states is only known in specific cases, namely when one
of the two states is pure~\cite{spedalieri_limit_2013} or for two
thermal states~\cite{scutaru}.

Here we solve this long-standing problem by deriving a computable
formula for the quantum fidelity between two arbitrary multimode
Gaussian states  which is simply expressed in terms of their
first- and second-order statistical moments. A key step for this
derivation relies on the adoption of an exponential Gibbs-like
representation for the Gaussian states, which has been used
recently to evaluate the fidelity between fermionic Gaussian
states~\cite{Fermion}, and which allows us to simplify many
calculations. We also provide a recipe for expressing the quantum
fidelity in terms of symplectic invariants, showing specific
examples with one, two and three modes. The new formula for the
fidelity allows us to easily derive the Bures metric for Gaussian
states, therefore generalizing quantum metrology to multimode
Gaussian resources. Similarly, we discuss how quantum hypothesis
testing can be extended beyond two-mode Gaussian states.

\section{Preliminary notions}

Consider $n$ bosonic modes described by quadrature operators $Q=(x_{1}%
,\dots,x_{n},p_{1},\dots,p_{n})^{T}$, satisfying the canonical
commutation relations~\cite{arvind_real_1995}
\begin{equation}
\lbrack Q,Q^{T}]=i\Omega,~~~~\Omega:=\begin{pmatrix}
0 & 1\\
-1 & 0
\end{pmatrix}
\otimes\openone,
\end{equation}
where $\openone$ is the $n\times n$ identity matrix. The
coordinate transformations $Q^{\prime}=SQ$ which preserve the
above commutation relations form the symplectic group, i.e. the
group of real matrices such that $S\Omega
S^{T}=\Omega$~\cite{Note1}.

Let us denote by $\rho$ an unnormalized density operator of the
$n$ bosonic modes. Its normalized version is denoted by
$\hat{\rho}=\rho/Z_{\rho}$, with $Z_{\rho}=\Tr\rho$ being the
normalization factor. For a Gaussian state~\cite{WeeRMP}, the
density operator $\hat{\rho}$ has a one-to-one correspondence with
the first- and second-order statistical moments of the state.
These are the mean value $u:=\langle
Q\rangle_{\hat{\rho}}=\mathrm{Tr}(Q\hat{\rho})\in \mathbb{R}^{2n}$
and the covariance matrix (CM) $V$, with generic element
\begin{equation}
V_{kl}=\frac{1}{2}\langle\{Q_{k}-u_{k},Q_{l}-u_{l}\}\rangle_{\hat{\rho}}~,
\end{equation}
where $\{,\}$ is the anticommutator. Equivalently, we may use the
following modified version of the CM
\begin{equation}
W:=-2Vi\Omega~.
\end{equation}

According to Williamson's theorem, there exists a symplectic
matrix $S$ such
that~\cite{WeeRMP}%
\begin{equation}
V=S(D\oplus D)S^{T},~~D=\diag(v_{1},\dots,v_{n}),
\end{equation}
where the symplectic eigenvalues satisfy $v_{k}\geq1/2$.
Correspondingly, the matrix $W$ transforms as $SWS^{-1}$ and its
standard eigenvalues are $\pm w_{k}$ where $w_{k}=2v_{k}\geq1$.

In Appendix~\ref{APPgen}, we show that an arbitrary multimode
Gaussian state with mean $u$ and CM $V$ can be written in the
exponential form
\begin{equation}
\rho=\exp\left[  -\frac{1}{2}(Q-u)^{T}G(Q-u)\right]
,~~Z_{\rho}=\det\left(
V+\frac{i\Omega}{2}\right)  ^{1/2}, \label{Gexpression}%
\end{equation}
where the Gibbs matrix $G$ is related to the CM by the formulae%
\begin{equation}
G=2i\Omega\,\coth^{-1}(2Vi\Omega),~~V=\frac{1}{2}\coth\left(
\frac{i\Omega
G}{2}\right)  i\Omega. \label{e.GtoV}%
\end{equation}
Equivalently, we may consider the following relations%
\begin{equation}
e^{i\Omega G}=\frac{W-\openone}{W+\openone},~~~~W=\frac{\openone+e^{i\Omega G}%
}{\openone-e^{i\Omega G}}~, \label{e.GtoW}%
\end{equation}
we use the notation $A/B:=AB^{-1}$ when $A$ and $B$ commute  --
see Appendix~\ref{appCOMP} for more details.
Although the matrix $G$ is singular for pure states (so one has to
deal carefully with this limit),  the introduction of the
representation in Eq.~\eqref{Gexpression} significantly simplifies
the calculations, and all the final formulae are valid in general,
i.e., for both mixed and pure states.

\section{Fidelity for multimode Gaussian states}

The quantum fidelity between two arbitrary quantum states, $\hat{\rho}%
_{1}=\rho_{1}/Z_{\rho_{1}}$ and
$\hat{\rho}_{2}=\rho_{2}/Z_{\rho_{2}}$, is
given by%
\begin{equation}
\mathcal{F}(\hat{\rho}_{1},\hat{\rho}_{2}):=\mathrm{Tr}\left(
\sqrt{\sqrt
{\hat{\rho}_{1}}\hat{\rho}_{2}\sqrt{\hat{\rho}_{1}}}\right) =\frac
{Z_{\sqrt{\rho_{\mathrm{tot}}}}}{\sqrt{Z_{\rho_{1}}Z_{\rho_{2}}}},
\label{fid2}%
\end{equation}
where
$\rho_{\mathrm{tot}}:=\sqrt{\rho_{1}}\rho_{2}\sqrt{\rho_{1}}$. We
consider two Gaussian states, $\hat{\rho }_{1}$ with CM\ $V_{1}$
and mean $u_{1}$, and $\hat{\rho}_{2}$ with CM\ $V_{2}$ and mean
$u_{2}$. The Gibbs matrices $G_1$ and $G_2$ are readily obtained
from Eqs.~\eqref{e.GtoV} and~\eqref{e.GtoW}. The advantage of the
Gibbs representation \eqref{Gexpression} for the calculation of
the fidelity is twofold: firstly, it makes the evaluation of the
operator square root in~Eq.\eqref{fid2} straightforward, and
secondly, one can use the algebra of quadratic operators
\cite{BalianNC} to find $\rho_{\mathrm{tot}}$ in a closed form.

As we show in Appendix~\ref{ProofAPP}, given two
generally-displaced Gaussian states, the formula for their quantum
fidelity can be directly expressed in terms of
$\delta_{u}:=u_{2}-u_{1}$ and their CMs, $V_{1}$ and $V_{2}$. In
fact, we find
\begin{equation}
\mathcal{F}(\hat{\rho}_{1},\hat{\rho}_{2})=
\mathcal{F}_0({V}_{1},{V}_{2})
\exp\left[  -\frac{1}{4}\delta
_{u}^{T}(V_{1}+V_{2})^{-1}\delta_{u}\right]  , \label{genFIDmain}%
\end{equation}
where the term $\mathcal{F}_0({V}_{1},{V}_{2}) $ depends only on
$V_1$ and $V_2$ and is easily computable from one of the two
auxiliary matrices
\begin{align}
V_{\mathrm{aux}}  &  = \Omega^{T}(V_{1}+V_{2})^{-1}\left(  \frac{\Omega}{4}%
+V_{2}\Omega V_{1}\right)  ,\label{Vauxi}\\
W_{\mathrm{aux}}  & := -2V_{\mathrm{aux}}i\Omega =
-(W_{1}+W_{2})^{-1}{(}\openone+W_{2}\,W_{1}).
\label{Wauxi}%
\end{align}
More precisely, we find%
\begin{align}
\mathcal{F}_0(V_{1},V_{2})&=
\frac{F_{\mathrm{tot}}}%
{\sqrt[4]{\det\left[  V_{1}+V_{2}\right]  }}~,
\\
F_{\mathrm{tot}}^{4}  &  =\det\left[  2\left(
\sqrt{\openone+\frac
{(V_{\mathrm{aux}}\Omega)^{-2}}{4}}+\openone\right)
V_{\mathrm{aux}}\right]
\label{Ftot1}\\
&  =\det\left[  \left(
\sqrt{\openone-W_{\mathrm{aux}}^{-2}}+\openone\right)
W_{\mathrm{aux}}i\Omega\right]  . \label{Ftot2}%
\end{align}
Note that the asymmetry of $V_{\mathrm{aux}}$ and
$W_{\mathrm{aux}}$ upon exchanging the two states is only apparent
and comes from the apparent asymmetry in the definition of
Eq.~\eqref{fid2}. One can check that the eigenvalues of
$V_{\rm{aux}}$ and $W_{\mathrm{aux}}$, and thus the determinants
in Eqs.~\eqref{Ftot1} and~\eqref{Ftot2}, are invariant under
exchange.



We remark that the formula of Eq.~(\ref{genFIDmain}) is valid for
arbitrary (generally-mixed) multimode Gaussian states with
arbitrary first- and second-order moments. In the specific case
where one of the states is pure (say $\rho_{1}$), we have
$V_{1}=\openone/2$ which implies $V_{\mathrm{aux}}=\openone/2$ and
$F_{\mathrm{tot}}=1$, therefore recovering the recent result of
Ref.~\cite{spedalieri_limit_2013} (in different
notation~\cite{Note2}).

\section{Fidelity in terms of symplectic invariants}

The fidelity can be expressed in terms of symplectic invariants
associated with the second-order moments of the Gaussian states.
Consider the notation with the $W$-matrices, so that
$F_{\mathrm{tot}}$\ is given by Eq.~(\ref{Ftot2}). The standard
eigenvalues of $W_{\mathrm{aux}}$ are $\pm w_{k}^{\mathrm{aux}}$,
where $w_{k}^{\mathrm{aux}}\geq 1$~\cite{WauxNOTE}. As a
consequence, we may write
\begin{equation}
F_{\mathrm{tot}}=\prod_{k=1}^{n}\left[  w_{k}^{\mathrm{aux}}+\sqrt
{(w_{k}^{\mathrm{aux}})^{2}-1}\right]  ^{1/2}. \label{FtotEIGEN}%
\end{equation}
Thus, the problem reduces to finding the eigenvalues of
$W_{\mathrm{aux}}$.

For this, let us consider the characteristic polynomial%
\begin{equation}
\chi(\lambda)=\det\left(  \lambda\openone-W_{\mathrm{aux}}\right)
, \label{e.charpol}
\end{equation}
which is clearly a symplectic invariant since $W_{\mathrm{aux}}$
transforms as $SW_{\mathrm{aux}}S^{-1}$ under symplectic
transformations. Using the identity $\det e^X=e^{\Tr X}$ and the
Cayley-Hamilton theorem \cite{MatrixBook}, we may write $\chi
(\lambda)$ as a polynomial function of
\begin{equation}
I_{2k}=\Tr(W_{\mathrm{aux}}^{2k}),~~\text{for~}k=1,...,n~,
\end{equation}
which are also symplectic invariants with $I_k>I_j$ for $k>j$.
Thus, for $n$ modes, we can compute the $n$ invariants $I_{2k}$
and subsequently solve the polynomial equation $\chi(\lambda)=0$,
whose roots are the eigenvalues $w_{k}^{\mathrm{aux}}$ to be used
in Eq.~(\ref{FtotEIGEN}).

Note that the invariants $I_{2k}$ can be connected with other
invariants. For instance, one can easily check that
\begin{equation}
\chi(0)=(-1)^{n}\frac{\Gamma}{\Delta},~~~~\chi(1)=(-1)^{n}\frac{\Lambda}{\Delta
}, \label{e.pmariaI}%
\end{equation}
where $\Delta:=\det(V_{1}+V_{2})$, $\Gamma:=2^{2n}\det(\Omega
V_{1}\Omega
V_{2}-\openone/4)$ and%
\begin{equation}
  \Lambda:=2^{2n}\det(V_{1}+i\Omega/2)\det(V_{2}+i\Omega/2) \label{e.MarianI}%
\end{equation}
are the invariants considered by Ref.~\cite{marian_uhlmann_2012}.
Using Eq.~(\ref{e.pmariaI}), one can easily express $I_{2}$\ and
$I_{4}$ in terms of $\Gamma$, $\Lambda$ and $\Delta$.

\section{Examples}

Let us show some examples with $n=1$, $2$ and $3$ modes. For
single-mode Gaussian states, we derive
$\chi(\lambda)=\lambda^{2}-I_{2}/2$, so that
$w^{\mathrm{aux}}=\sqrt{I_{2}/2}$. Equivalently, we may compute $I_{2}%
/2=1+\Lambda/\Delta$ so that we retrieve the known
result~\cite{fidG2,fidG3,fidG4}%
\begin{equation}
  \mathcal{F}^2_0({V}_{1},{V}_{2})
=\frac{1}{\sqrt
  {\Delta+\Lambda}-\sqrt{\Lambda}}.
\end{equation}

For two-mode Gaussian states, we derive $\chi(\lambda)=(I_{2}^{2}%
-2I_{4}-4{I_{2}}\lambda^{2}+8\lambda^{4})/8$ with solutions
\begin{equation}
w_{\pm}^{\mathrm{aux}}=\frac{1}{2}\sqrt{I_{2}\pm\sqrt{4I_{4}-I_{2}^{2}}}.
\end{equation}
Once plugged into Eq.~(\ref{FtotEIGEN}), we have the fidelity in
terms of $I_{2}$ and $I_{4}$. The latter invariants can then be
expressed in terms of $\Gamma/\Delta$ and $\Lambda/\Delta$, so
that we retrieve the known result~\cite{marian_uhlmann_2012}
\begin{equation}
  \mathcal{F}^2_0({V}_{1},{V}_{2})
=\frac{1}{\sqrt{\Gamma }+\sqrt{\Lambda}-\sqrt{\left(
\sqrt{\Gamma}+\sqrt{\Lambda}\right) ^{2}-\Delta}}.
\end{equation}

For three-mode Gaussian states, the characteristic polynomial may
be written
as $\chi=t^{3}+pt+q$, where%
\begin{equation}
t=\lambda^{2}-I_{2}/6,~~p=\frac{I_{2}^{2}}{24}-\frac{I_{4}}{4},~~q=-\frac
{I_{2}^{3}}{108}+\frac{I_{2}I_{4}}{12}-\frac{I_{6}}{6}.
\label{e.I3}
\end{equation}
The solutions of the characteristic equation $\chi=0$ are real
(see Appendix~\ref{RealSolu}) and given by
\begin{equation}
w_{k}^{\mathrm{aux}}=\sqrt{\frac{I_{2}}{6}+2\sqrt{-\frac{p}{3}}\cos\left[
\frac{\theta-2\pi(k-1)}{3}\right]  }, \label{e.waux3}%
\end{equation}
where $\theta:=\arccos\left[ 3\sqrt{3}q(2p\sqrt{-p})^{-1}\right] $
and $k=1,2,3$ (in particular, note that
$w_{k}^{\mathrm{aux}}=\sqrt{{I_{2}}/{6}}$ for $p=0$). To the best
of our knowledge, Eqs.~\eqref{e.I3} and~\eqref{e.waux3}, together
with Eqs.~\eqref{genFIDmain} and~\eqref{FtotEIGEN}, provide the
first expression for the quantum fidelity between two arbitrary
three-mode Gaussian states.

\section{Immediate implications}

\subsection{Geometry of Gaussian states}

Once the quantum fidelity is expressed in terms of the first two
statistical moments, we can easily compute the Bures distance
between two arbitrary multimode Gaussian states, $\hat{\rho}_{1}$
and $\hat{\rho}_{2}$, which is given by
\begin{equation}
D_{B}(\hat{\rho}_{1},\hat{\rho}_{2})=2\left[  1-\mathcal{F}(\hat{\rho}%
_{1},\hat{\rho}_{2})\right]  .
\end{equation}
Form this expression we can derive the Bures metric by expanding
the fidelity. In fact, let us consider two infinitesimally-close
Gaussian states $\hat{\rho }_{1}=\hat{\rho}$, with statistical
moments $u$ and $V$, and $\hat{\rho}_{2}=\hat{\rho }+d\hat{\rho}$,
with statistical moments $u+du$ and $V+dV$. Then, the Bures metric
is given by
\begin{equation}
ds^{2} =2[
1-\mathcal{F}(\hat{\rho},\hat{\rho}+d\hat{\rho})]=\frac{du^{T}V^{-1}du}{4}
+\frac{\delta}{8}~, \label{BuresM}
\end{equation}
where $\delta:=4\Tr[dV
(4\mathcal{L}_{V}+\mathcal{L}_{\Omega})^{-1}dV]$,
$\mathcal{L}_{A}X:=AXA$, and the inverse of the superoperator
$4\mathcal{L}_{V}+\mathcal{L}_{\Omega}$ refers to the
pseudo-inverse~\cite{MatrixBook} (see Appendix~\ref{BuresAPP} for
the proof). Note that a result equivalent to Eq.~(\ref{BuresM})
has been derived in Ref.~\cite{Monras} using a different method
based on the computation of the symmetric logarithmic derivative.

Numerically, the easiest way of evaluating the inverse of the
superoperator in $\delta$ is using the $W$-matrices and performing
the calculations in the basis in which $W$ is diagonal. In the
basis where $W$ is diagonal, then
\begin{equation}
\delta=\sum_{ij}\frac{dW_{ij}dW_{ji}}{w_{i}w_{j}-1} ~,
\label{BuresBasis}
\end{equation}
and the sum is taken over the elements such that $w_iw_j\neq1$.
For pure states, we simply have $\delta_{\mathrm{pure}}
=\Tr(V^{-1}\,dV\,V^{-1}\,dV)$.

\subsection{Multimode quantum metrology}

Let us consider a real parameter $\theta$ which is encoded in a
multimode Gaussian state $\hat{\rho}_{\theta}$. To estimate
$\theta$ with high precision, it is necessary to distinguish the
two infinitesimally-close states $\hat{\rho}_{\theta}$ and
$\hat{\rho}_{\theta+d\theta}$\ for an infinitesimal change
$d\theta$. Assume that $N$ copies of the state $\hat
{\rho}_{\theta}$ are available to an observer, who performs $N$
independent measurements to obtain an unbiased estimator
$\tilde{\theta}$ for parameter $\theta$. Then, the mean-square
error affecting the parameter estimation
\textrm{Var}$(\theta):=\langle(\tilde{\theta}-\theta)^{2}\rangle$
satisfies the quantum Cramer-Rao (QCR) bound
$\mathrm{Var}(\theta)\geq\lbrack NH(\theta)]^{-1}$, where
$H(\theta)$ is the quantum Fisher information
(QFI)~\cite{SamMETRO}. The latter can be computed from the
fidelity as
\begin{equation}
H(\theta)=\frac{8\left[  1-\mathcal{F}(\hat{\rho}_{\theta},\hat{\rho}%
_{\theta+d\theta})\right]  }{d\theta^{2}}~. \label{QFImain}%
\end{equation}
Thus, for any parametrization of the Gaussian states, we can
easily compute the fidelity
$\mathcal{F}(\hat{\rho}_{\theta},\hat{\rho}_{\theta+d\theta})$
using Eq.~(\ref{genFIDmain}) and, therefore, the QFI in
Eq.~(\ref{QFImain}).

More generally, suppose that the Gaussian state is labelled by a
vectorial parameter with $m$ real components, i.e.,
$\theta=\{\theta_{i}\}$ for $i=1,\ldots,m$. In this case, the
performance of the parameter estimation is expressed by the
classical covariance matrix \textrm{Cov}$_{ij}(\theta
):=\langle\tilde{\theta}_{i}\theta_{j}\rangle-\langle\tilde{\theta}_{i}%
\rangle\langle\theta_{j}\rangle$, which satisfies the matrix
version of the QCR bound~\cite{QCRbound,Paris}
$\mathrm{Cov}(\theta)\geq\lbrack NH(\theta)]^{-1}$. Here the QFI
is a matrix with elements $H_{ij}(\theta)$, which can be evaluated
from the Bures metric. In fact, for any
parametrization, we may write Eq.~(\ref{BuresM}) as $ds^{2}=g_{ij}%
(\theta)d\theta_{i}d\theta_{j}$ and show that
$H_{ij}(\theta)=4g_{ij}(\theta)$.

\subsection{Multimode quantum hypothesis testing}

An efficient computation of the quantum fidelity is crucial for
solving problems of binary quantum hypothesis
testing~\cite{QHT,QHT2} with multimode Gaussian states. These
problems may occur in the basic scenario of quantum state
discrimination, where two Gaussian states must be optimally
distinguished, or in the setting of quantum channel
discrimination, where two Gaussian channels must be distinguished
by assuming Gaussian sources and input energy constraints. In
particular, the latter formulation is very important in a variety
of quantum technology protocols, such as remote quantum sensing of
targets, i.e., quantum illumination~\cite{Qill,Qill2,Qill3}, and
quantum reading of classical data from optical
memories~\cite{Qread,Nair11,Hirota11, Bisio11, Arno11,Qread3}.

Consider $N$ copies of two multimode Gaussian states,
$\hat{\rho}_{1}^{\otimes N}$ and $\hat{\rho}_{2}^{\otimes N}$,
with the same a priori probability. The minimum error probability
$p_{\text{err}}(N)$ in their statistical discrimination is
provided by the Helstrom bound~\cite{Helstrom}, which is typically
hard to compute for mixed states. For this reason, one resorts to
other computable bounds, such as the quantum Chernoff
bound~\cite{QCbound,QCB,MinkoPRA} or fidelity-based
bounds~\cite{Fuchs,FuchsThesis,MinkoPRA}. Thanks to our result the
latter are now the simplest to compute.

For any number of copies $N$, we may write
\begin{equation}
\frac{1-\sqrt{1-\left[
\mathcal{F}(\hat{\rho}_{1},\hat{\rho}_{2})\right]
^{2N}}}{2}\leq p_{\text{err}}(N)\leq\frac{\left[  \mathcal{F}(\hat{\rho}_{1}%
,\hat{\rho}_{2})\right]  ^{N}}{2}~.\label{errPROB}%
\end{equation}
In particular, the lower bound in Eq.~(\ref{errPROB}) is the
tightest known. Note that Eq.~(\ref{errPROB}) can be derived by
using the known result for single copy ($N=1$)~\cite{Fuchs} and
then applying the multiplicative property of the fidelity under
tensor products of density operators, so that
$\mathcal{F}(\hat{\rho}_{1}^{\otimes N},\hat{\rho}_{2}^{\otimes
N})=\mathcal{F}(\hat{\rho}_{1},\hat{\rho}_{2})^N$.

The computation of the quantum fidelity is also important for
asymmetric quantum hypothesis testing where the two quantum
hypotheses have unbalanced Bayesian costs~\cite{QHB}. In this
context, the quantum fidelity can be used to estimate the quantum
Hoeffding bound~\cite{QHBgae} which quantifies the optimal
error-exponent associated with the rate of false negatives.

\section{Conclusions}

In this work we have solved a long-standing open problem in
continuous variable quantum information by deriving a simple
computable formula for the quantum fidelity between two arbitrary
multimode Gaussian states. Our main formula is expressed in terms
of the statistical moments of the Gaussian states, but another
formulation is also given in terms of suitable symplectic
invariants. By using our formula, one can extend the study of
quantum teleportation, cloning, quantum metrology and hypothesis
testing well beyond the standard case of two-mode Gaussian states
to consider multimode Gaussian resources, with unexplored
implications for all these basic quantum information protocols.

\section*{Acknowledgments}
L.B. is currently supported by the European Research Council under
the European Union's Seventh Framework Programme (FP/2007-2013) /
ERC Grant Agreement n. 308253. S.P. has been supported by the
Leverhulme Trust (`qBIO' fellowship) and EPSRC via `qDATA' (Grant
No. EP/L011298/1) and the UK Quantum Communications Hub (Grant No.
EP/M013472/1).


\appendix

\part*{~~~~~~~~~~~~~Appendices}

\section{Exponential formula for Gaussian states\label{APPgen}}

Here we show the formulae in Eqs.~(\ref{Gexpression})-(\ref{e.GtoW}). The
first step is to introduce the symplectic action of a real function $f$ on a
CM and how it can be computed in terms of standard matrix functions when $f$
is odd. After this preliminary step, we start by noting that, for thermal
states (having $V=D\oplus D$), we can easily write Eq.~(\ref{Gexpression})
with $u=0$ and%
\begin{equation}
G=g(D)\oplus g(D),~~g(v)=2\coth^{-1}(2v). \label{gsmall}%
\end{equation}
Then, we generalize the formula to zero-mean Gaussian states with arbitrary
CMs by noting that $\Omega G\Omega$ transforms as $V$ under symplectic
coordinate transformations $Q^{\prime}=SQ$. This property allows us to use the
symplectic action $g_{\ast}(v)$ which leads to Eq.~(\ref{e.GtoV}). Finally, we
include displacements to extend the result to arbitrary mean values and we
compute the normalization factor.

\subsection{Symplectic action and its computation\label{app1}}

Then, let $f:\mathbb{R}\rightarrow\mathbb{R}$ be a function. The symplectic
action $f_{\ast}$ on the CM\ $V$\ is defined by~\cite{spedalieri_limit_2013}
\begin{equation}
f_{\ast}(V)=S[f(D)\oplus f(D)]S^{T}, \label{SA}%
\end{equation}
where $f(D)=\diag[f(v_{1}),f(v_{2}),\dots,f(v_{n})]$ acts as a standard matrix
function. Here we prove that, if $f$ is an odd function $f(-x)=-f(x)$, then
\begin{equation}
f_{\ast}(V)=f(Vi\Omega)i\Omega. \label{simpleSA}%
\end{equation}

Let us start by proving that Eq.~(\ref{simpleSA}) satisfies the identity
\begin{equation}
f_{\ast}(SVS^{T})=Sf_{\ast}(V)S^{T}~. \label{SA_identity}%
\end{equation}
In fact, we have
\begin{align*}
f_{\ast}(SVS^{T})  &  =f(SVS^{T}i\Omega)i\Omega=f(SVi\Omega S^{-1})i\Omega\\
&  =Sf(Vi\Omega)S^{-1}i\Omega=Sf(Vi\Omega)i\Omega S^{T}\\
&  =Sf_{\ast}(V)S^{T}~,
\end{align*}
where we use the basic property $f(SVS^{-1})=Sf(V)S^{-1}$.

Because of Eq.~(\ref{SA_identity}), without loss of generality, we can focus
on the case where $V$ is in diagonal Williamson form, i.e.,
\[
V=D\oplus D,~~D=\diag(v_{1},v_{2},\dots,v_{n}),
\]
and we assume that $v_{i}\neq v_{j}$ for $i\neq j$. One can easily check that
the matrix
\[
\tilde{V}=(D\oplus D)i\Omega
\]
is Hermitian, so it can be cast into the diagonal form by a unitary matrix
$U$. It turns out that $U$ is independent on $v_{i}$ and
\begin{equation}
\tilde{V}=U^{\dagger}(D\oplus-D)U, \label{structure}%
\end{equation}
with eigenvalues $\pm v_{i}$. If $f$ is an odd function, then%
\[
f(\tilde{V})=U^{\dagger}[f(D)\oplus f(-D)]U=U^{\dagger}[f(D)\oplus-f(D)]U.
\]
The latter matrix has the same structure of $\tilde{V}$ in
Eq.~(\ref{structure}). Because $U$ is independent on the diagonal elements,
then%
\[
f(\tilde{V})=[f(D)\oplus f(D)]i\Omega~,
\]
which gives%
\[
f[(D\oplus D)i\Omega]i\Omega=f(D)\oplus f(D).
\]
This is Eq.~(\ref{simpleSA}) up to a symplectic transformation $S$.

\subsection{Proof of the exponential formula\label{appEXP}}

Let us now show that the Gibbs exponential formula of Eq.~(\ref{Gexpression})
can describe an arbitrary Gaussian state (not just a thermal state). We start
by considering a single-mode thermal state $\rho=e^{-ga^{\dagger}a}$. In this
case, we can write%
\begin{equation}
\tilde{Z}_{\rho}=\frac{1}{1-e^{-g}},~~\langle{a^{\dagger}a}\rangle=-\frac
{1}{\tilde{Z}}\frac{\partial\tilde{Z}}{\partial g}=\frac{1}{e^{g}-1}.
\label{eqq1}%
\end{equation}
In our notation, $a=(x+ip)/\sqrt{2}$ so that $a^{\dagger}a=\frac{x^{2}+p^{2}%
}{2}-\frac{1}{2}$ and
\begin{equation}
v(g):=\langle{x^{2}}\rangle=\langle{p^{2}}\rangle=\langle{a^{\dagger}a}%
\rangle+1/2. \label{eqq2}%
\end{equation}
Therefore, from Eqs.~(\ref{eqq1}) and~(\ref{eqq2}), we derive
\begin{equation}
v(g)=\frac{1}{2}\coth\frac{g}{2}. \label{invFUN}%
\end{equation}
In terms of the quadratures, the thermal state reads%
\begin{equation}
\rho=e^{-\frac{g}{2}\left(  x^{2}+p^{2}\right)  }, \label{Gquadr}%
\end{equation}
and its normalization is given by%
\begin{equation}
Z_{\rho}=\tilde{Z}_{\rho}e^{-\frac{g}{2}}=\frac{1}{e^{\frac{g}{2}}%
-e^{-\frac{g}{2}}}:=z(g). \label{e.singlexp}%
\end{equation}
Note that the purity is given by
\[
\Tr\hat{\rho}^{2}=Z_{\rho^{2}}/Z_{\rho}^{2}=z(2g)z^{-2}(g)=\tanh(g/2)=\frac
{1}{2}v(g)^{-1},
\]
so that the vacuum corresponds to $g\rightarrow\infty$ or $v\rightarrow1/2$.

The previous representation of Eq.~(\ref{Gquadr}) can be generalized to a
multimode thermal state of $n\geq1$ bosonic modes. This state has its
CM\ already in the diagonal Williamson form
\[
V=D\oplus D,~~D=\mathrm{diag}(v_{1},\ldots,v_{n}).
\]
Thanks to the tensor product structure, we can write%
\begin{equation}
\rho=e^{-\frac{1}{2}Q^{T}GQ}~. \label{appTOT}%
\end{equation}
Here $G:=\diag(g_{1},\dots,g_{n};g_{1},\dots,g_{n})$, where the diagonal
elements are given by $g_{i}=g(v_{i})$, where%
\begin{equation}
g(v)=2\coth^{-1}(2v) \label{gAPP}%
\end{equation}
is the inverse of the function in Eq.~(\ref{invFUN}). Compactly, we set%
\[
G=g(D)\oplus g(D).
\]

Now, we study how $G$ and $V$ transform under coordinate transformations
$Q^{\prime}=SQ$. We have $V^{\prime}=SVS^{T}$ and%
\begin{equation}
G^{\prime}=S^{-T}GS^{-1}=\Omega S\Omega G\Omega S^{T}\Omega,
\label{e.coordtrG}%
\end{equation}
where Eq.~\eqref{e.coordtrG} comes from imposing $Q^{T}GQ=Q^{\prime}{}%
^{T}G^{\prime}Q^{\prime}$ in Eq.~(\ref{appTOT}). From Eq.~\eqref{e.coordtrG},
we see that
\[
(\Omega G\Omega)\rightarrow S(\Omega G\Omega)S^{T},
\]
i.e., matrices $V$ and $\Omega G\Omega$ transform in the same way under
symplectic coordinate transformations. As a result, they can be related by the
symplectic action of the function in Eq.~(\ref{gAPP}).

In fact, for thermal states, we may write
\[
V=D\oplus D,~~(\Omega G\Omega)=-g(D\oplus D).
\]
Then, for an arbitrary symplectic transformation $S$, we have%
\[%
\begin{array}
[c]{ccc}%
\text{Thermal} & ~~~~~~ & \text{Arbitrary}%
~~~~~~~~~~~~~~~~~~~~~~~~~~~~~~~~~~~~~~~~~~~~~\\
D\oplus D & \overset{S}{\rightarrow} & V=S(D\oplus D)S^{T}%
~~~~~~~~~~~~~~~~~~~~~~~~~~~~~~~~\\
-g(D\oplus D) & \overset{S}{\rightarrow} & \Omega G\Omega=S\left[  -g(D\oplus
D)\right]  S^{T}=-g_{\ast}(V).
\end{array}
\]
Thus, using the symplectic action $g_{\ast}$, defined from Eq.~(\ref{gAPP}),
and its inverse $\nu_{\ast}$, defined from Eq.~(\ref{invFUN}),\ we can derive
the relations%
\[
G=-2\Omega\,\coth_{\ast}^{-1}(2V)\Omega=2i\Omega\,\coth^{-1}(2Vi\Omega),
\]
and%
\[
V=-\frac{1}{2}\coth_{\ast}\left(  \frac{\Omega G\Omega}{2}\right)  =\frac
{1}{2}\coth\left(  \frac{i\Omega G}{2}\right)  i\Omega,
\]
where we also exploit Eq.~(\ref{simpleSA}). These formulae correspond to those
in Eq.~(\ref{e.GtoV}) given in the main text. The additional formula in
Eq.~(\ref{e.GtoW}) is obtained by considering that $W=-2Vi\Omega$.

\subsubsection{Extension to non-zero mean}

The next step is to include the presence of a generally non-zero mean value in
the exponential expression of Eq.~(\ref{appTOT}). For an arbitrary
$u\in\mathbb{R}^{2n}$, consider the displacement operator%
\[
D(u)=e^{u^{T}i\Omega Q}=e^{-iQ^{T}\Omega u},
\]
which satisfies $D(u)^{\dagger}=D(-u)$ and $D(u)QD(u)^{\dagger}=Q+u$. By
applying this operator to Eq.~(\ref{appTOT}), we can generate an arbitrary
Gaussian state with non-zero mean%
\[
\rho=D(-u)e^{-\frac{1}{2}Q^{T}GQ}D(u)=e^{-\frac{1}{2}(Q-u)^{T}G(Q-u)}.
\]
This is easy to double check. Let us set
\[
\rho=D(-u)\rho_{G}D(u),~\rho_{G}:=e^{-\frac{1}{2}Q^{T}GQ}.
\]
First note that $Z_{\rho}=Z_{\rho_{G}}$. Then, we can verify that
\begin{align*}
\Tr[Q \,\frac{e^{-\frac12 (Q-u)^T\,G\,(Q-u)}}{Z_\rho}]  &
=\Tr[Q \,D(-u) \frac{e^{-\frac12 Q^T\,G\,Q}}{Z_\rho}D(u)]=\\
\Tr[D(u)\,Q \,D(-u) \frac{e^{-\frac12 Q^T\,G\,Q}}{Z_\rho}]  &
=\Tr[(Q +u)\, \frac{e^{-\frac12 Q^T\,G\,Q}}{Z_{\rho_G}}]=u,
\end{align*}
i.e. $\langle Q\rangle_{\hat{\rho}}=u$. Similarly, $V_{ij}=\frac{1}{2}%
\langle\{Q_{i}-u_{i},Q_{j}-u_{j}\}\rangle_{\hat{\rho}}$.

\subsubsection{Normalization factor}

The trace of an unnormalized Gaussian state $\rho$ is written in
Eq.~(\ref{eqq1}) via the function $z(g)=1/(e^{g/2}-e^{-g/2})$ defined in
Eq.~\eqref{e.singlexp}. When $G$ is diagonal (i.e. $V$ is diagonal) then
\begin{equation}
Z_{\rho}=\prod_{j}z(g_{j})~. \label{Zdiag}%
\end{equation}
Now we write Eq.~(\ref{Zdiag}) in a coordinate independent form. A generic $G$
can be obtained from a diagonal $G$ via a symplectic coordinate
transformation, because of the property~(\ref{SA}) of the symplectic action,
and because $\det S=1$, one has
\begin{align}
Z_{\rho}  &  =\sqrt{\det[z_{\ast}(G)]}=\det[z(G\,i\Omega)\,i\Omega]^{1/2}%
=\det[z(i\Omega\,G)\,i\Omega]^{1/2}\nonumber\\
&  =\det[\left(  e^{i\Omega G/2}-e^{-i\Omega G/2}\right)  \,i\Omega]^{-1/2}~.
\label{e.Zg}%
\end{align}
Moreover, $z(g(v))=\sqrt{v^{2}-\frac{1}{4}}$. It is simple to prove that
\begin{align}
Z_{\rho}  &  =\prod_{j}z(g(v_{j}))=\prod_{j}\sqrt{v_{j}^{2}-\frac{1}{4}}\\
&  =\det[V_{\mathrm{diag}}+i\Omega/2]^{1/2} \label{e.Zvdiag}%
\end{align}
where $V_{\mathrm{diag}}=\diag(v_{1},\dots,v_{n};v_{1},\dots,v_{n})$. Since a
general $V$ can be written as $V=S\,V_{\mathrm{diag}}\,S^{T}$ and $\det S=1$,
then
\[
Z_{\rho}=\det\left(  V+\frac{i\Omega}{2}\right)  ^{1/2}~,
\]
where we used the fact that $S\Omega S^{T}=\Omega$. By replacing
$W=-2Vi\Omega$, we also get%
\[
Z_{\rho}=\det\left(  \frac{\openone-W}{2i\Omega}\right)  ^{1/2}.
\]

\section{Computations with Gaussian states\label{appCOMP}}

\subsection{Product of two Gaussian states with zero mean}

Although the product of two Gaussian states can be readily evaluated thanks
to the result of \cite{BalianNC}, in this section we provide a self-consistent
proof.

By using the Baker-Campbell-Hausdorff identity, we can write the product of
two zero-mean Gaussian states as%
\begin{equation}
e^{-\frac{1}{2}Q^{T}GQ}\,e^{-\frac{1}{2}Q^{T}G^{\prime}Q}=e^{-\frac{1}{2}%
Q^{T}G^{\prime\prime}Q}~. \label{e.bchrho2}%
\end{equation}
The above identity is a consequence of the algebra
\begin{align}
\left[  -\frac{Q^{T}\tilde{G}Q}{2},-\frac{Q^{T}\tilde{G}^{\prime}Q}{2}\right]
&  =\frac{i}{2}Q^{T}\left(  \tilde{G}\Omega\tilde{G}^{\prime}-\tilde
{G}^{\prime}\Omega\tilde{G}\right)  Q\nonumber\\
&  =-\frac{Q^{T}\tilde{G}^{\prime\prime}Q}{2}, \label{e.algebra}%
\end{align}
where%
\[
\tilde{G}=-i\Omega\tilde{J},~\tilde{G}^{\prime}=-i\Omega\tilde{J}^{\prime
},~\tilde{G}^{\prime\prime}=-i\Omega\tilde{J}^{\prime\prime}%
\]
and $J^{\prime\prime}=[J,J^{\prime}]$. Because of the above identity, we can
write the Eq.~\eqref{e.bchrho2} with $e^{J^{\prime\prime}}=e^{J}e^{J^{\prime}%
}$, namely
\begin{equation}
e^{-i\Omega\,G^{\prime\prime}}=e^{-i\Omega\,G}e^{-i\Omega\,G^{\prime}}.
\label{compRULE}%
\end{equation}

Now we can express the composition rule of Eq.~(\ref{compRULE}) in terms of
the CMs $V$ and $V^{\prime}$ of the two states $\rho$ and $\rho^{\prime}$.
From Eq.~\eqref{e.GtoV}, we have%
\[
V=\frac{1}{2}\frac{e^{i\Omega G}+\openone}{e^{i\Omega G}-\openone}%
i\Omega,~e^{i\Omega G}=\frac{-2Vi\Omega-\openone}{-2Vi\Omega+\openone}.
\]
In terms of $W=-2Vi\Omega$, $W^{\prime}$ and $W^{\prime\prime}$, we may write%
\begin{align*}
W^{\prime\prime}  &  =-\frac{e^{i\Omega G^{\prime}}e^{i\Omega G}%
+\openone}{e^{i\Omega G^{\prime}}e^{i\Omega G}-\openone}=-\frac{\frac
{W^{\prime}-\openone}{W^{\prime}+\openone}+\frac{W+\openone}{W-\openone}%
}{\frac{W^{\prime}-\openone}{W^{\prime}+\openone}-\frac{W+\openone}%
{W-\openone}}\\
&  =-\frac{\openone-2(W^{\prime}+\openone)^{-1}+\openone+2(W-\openone)^{-1}%
}{\openone-2(W^{\prime}+\openone)^{-1}-\openone-2(W-\openone)^{-1}}\\
&  =\frac{\openone-(W^{\prime}+\openone)^{-1}+(W-\openone)^{-1}}{(W^{\prime
}+\openone)^{-1}+(W-\openone)^{-1}}\\
&  =\openone+\frac{\openone-2(W^{\prime}+\openone)^{-1}}{(W^{\prime
}+\openone)^{-1}+(W-\openone)^{-1}}.
\end{align*}
In the above equations, we fix the notation $\frac{A}{B}=AB^{-1}$ when
$[A,B]\neq 0$.
Using the Woodbury identity \cite{MatrixBook}
\[
  (A+B)^{-1}=A^{-1}-A^{-1}(A^{-1}+B^{-1})^{-1}A^{-1}%
\]
we derive
\[
W^{\prime\prime}=\openone+(W^{\prime}+\openone-2\openone)[\openone-(W^{\prime
}+W)^{-1}(W^{\prime}+\openone)]~.
\]
Then, using another straightforward matrix equation
\begin{equation}
(A+B)^{-1}A=\openone-(A+B)^{-1}B,
  \label{dyson}
\end{equation}
with $A=W^{\prime}+\openone$ and $B=W-\openone$, we find
\begin{equation}
W^{\prime\prime}=\openone+(W^{\prime}-\openone)(W^{\prime}+W)^{-1}%
(W-\openone). \label{Wsec}%
\end{equation}
Therefore
\begin{equation}
V^{\prime\prime}=-\frac{i\Omega}{2}+\left(  V^{\prime}+\frac{i\Omega}%
{2}\right)  (V^{\prime}+V)^{-1}\left(  V+\frac{i\Omega}{2}\right)  .
\label{Vsec}%
\end{equation}

Note that the squared of a Gaussian state $\rho^{2}$ has $G^{(2)}=2G$ and its
CM can be computed directly from the previous Eqs.~(\ref{Wsec})
and~(\ref{Vsec}) by setting $W=W^{\prime}$ and $V=V^{\prime}$. It is easy to
check that we get%
\[
V^{(2)}=\frac{1}{2}\left(  V-\frac{\Omega V^{-1}\Omega}{4}\right)
,~W^{(2)}=\frac{1}{2}\left(  W+W^{-1}\right)  .
\]

\subsection{Square root of Gaussian states}

Given a Gaussian state $\rho$, its square-root $\sqrt{\rho}$ is a state with
$G\rightarrow G/2$. The CM $V_{\mathrm{sq}}$ of $\sqrt{\rho}$ can be written
in terms of the CM $V$ of $\rho$ by concatenating functions
\begin{equation}
v_{\mathrm{sq}}(v):= v(g(v)/2)=\left(
\sqrt{1-\frac{1}{4v^{2}}}+1\right)
v. \label{corrEQ}%
\end{equation}
Notice that, because $v\geq1/2$ one might be tempted to simplify
$v_{\mathrm{sq}}(v)$ into the expression $x+\sqrt{4x^{2}-1}/2$. However, the
latter function is not odd, so it produces wrong results when it is used for
the symplectic action. Eq.~(\ref{corrEQ}) is the correct one. When $V$ is
Williamson-diagonal, so it is $V_{\mathrm{sq}}$ and the diagonal elements are
given by $[V_{\mathrm{sq}}]_{ii}=v_{\mathrm{sq}}(v_{i})$. Since $V$ and
$V_{\mathrm{sq}}$ transform in the same way under symplectic transformations,
for any general (non-diagonal) $V$, the relation between $V$ and
$V_{\mathrm{sq}}$ can be obtained with the symplectic action%
\[
V_{\mathrm{sq}}=v_{\mathrm{sq,\ast}}(V)=\left(  \sqrt{\openone+\frac
{(V\,\Omega)^{-2}}{4}}+\openone\right)  V~.
\]
By replacing $W=-2Vi\Omega$, we finally derive%
\begin{equation}
W_{\mathrm{sq}}=\left(  \sqrt{\openone-W^{-2}}+\openone\right)  W.
\label{sqCM}%
\end{equation}

\subsection{Extending the product formula to Gaussian states with non-zero
mean}

When an operator linear in terms of $Q$ is introduced, the algebra in
Eq.~\eqref{e.algebra} has to be extended. It turns out that
\begin{align}
\left[  -\frac{1}{2}Q^{T}GQ,Q\right]   &  =i\Omega GQ~,\nonumber\\
\lbrack u^{T}Q,v^{T}Q]  &  =u^{T}i\Omega v~. \label{e.algebrad}%
\end{align}
Therefore, $D(u)QD(u)^{\dagger}=Q+u$, and using Eqs.~\eqref{e.algebra}
and~\eqref{e.algebrad}, we may write the identities
\begin{align}
D(u)D(v)  &  =D(u+v)e^{-\frac{1}{2}u^{T}i\Omega v},\nonumber\\
e^{-\frac{1}{2}Q^{T}GQ}Qe^{\frac{1}{2}Q^{T}GQ}  &  =e^{i\Omega G}Q.
\label{e.iddisplaced}%
\end{align}

\subsection{Decomposition of displaced Gaussian states}

Using the previous identities we may write%
\begin{align*}
\rho &  =e^{-\frac{1}{2}(Q-u)^{T}G(Q-u)}=e^{-u^{T}i\Omega Q}e^{-\frac{1}%
{2}Q^{T}GQ}e^{iu^{T}i\Omega Q}\\
&  =e^{-u^{T}i\Omega Q}e^{u^{T}i\Omega e^{i\Omega G}Q}e^{-\frac{1}{2}Q^{T}%
GQ}\\
&  =e^{-u^{T}i\Omega Q}e^{u^{T}e^{Gi\Omega}i\Omega Q}e^{-\frac{1}{2}Q^{T}GQ}\\
&  =e^{-u^{T}i\Omega Q}e^{(e^{-i\Omega G}u)^{T}i\Omega Q}e^{-\frac{1}{2}%
Q^{T}GQ}\\
&  =e^{\frac{1}{2}u^{T}i\Omega e^{-i\Omega G}u}e^{(e^{-i\Omega G}%
u-u)^{T}i\Omega Q}e^{-\frac{1}{2}Q^{T}GQ}.
\end{align*}
Let $\ell=e^{-i\Omega G}u-u$, i.e.
\[
u=(e^{-i\Omega G}-\openone)^{-1}\ell.
\]
Note that
\begin{align}
u^{T}i\Omega e^{-i\Omega G}u  &  =\Re\lbrack u^{T}i\Omega e^{-i\Omega
G}u]\nonumber\\
&  =\frac{1}{2}u^{T}i\Omega(e^{-i\Omega G}-e^{i\Omega G})u~.
\label{e.alternativeNorm}%
\end{align}
Then, using the above result
\begin{align}
e^{\ell^{T}i\Omega Q}e^{-\frac{1}{2}Q^{T}GQ}  &  =e^{-\frac{1}{2}%
(Q-u)^{T}G(Q-u)}e^{-\frac{1}{2}u^{T}i\Omega e^{-i\Omega G}u}\nonumber\\
&  =\rho(G,u)\,e^{-K}, \label{e.complinear}%
\end{align}
where
\begin{align*}
K  &  =\frac{1}{4}u^{T}i\Omega(e^{-i\Omega G}-e^{i\Omega G})u\\
&  =\frac{1}{4}\ell^{T}(e^{Gi\Omega}-\openone)^{-1}i\Omega(e^{-i\Omega
G}-e^{i\Omega G})(e^{-i\Omega G}-\openone)^{-1}\ell\\
&  =\frac{1}{4}\ell^{T}i\Omega(e^{i\Omega G}-\openone)^{-1}(e^{-i\Omega
G}-e^{i\Omega G})(e^{-i\Omega G}-\openone)^{-1}\ell\\
&  =\frac{1}{4}\ell^{T}i\Omega\left(  (e^{i\Omega G}-\openone)^{-1}%
-(e^{-i\Omega G}-\openone)^{-1}\right)  \ell\\
&  =\frac{1}{4}\ell^{T}i\Omega\left(  -\frac{W+\openone}{2}-\frac
{W-\openone}{2}\right)  \ell\\
&  =-\frac{1}{4}\ell^{T}i\Omega W\ell.
\end{align*}

\section{Proof of Eq.~(\ref{genFIDmain})\label{ProofAPP}}

We start by considering the undisplaced case where $u_1=u_2=0$. This assumption
will be relaxed in Appendix~\ref{displAPP}.

The total state $\rho_{\mathrm{tot}}:=\sqrt{\rho_{1}}\rho_{2}\sqrt{\rho_{1}}$
has CM $V_{\mathrm{tot}}$ ($W_{\mathrm{tot}}$)\ and its Gibbs matrix
$G_{\mathrm{tot}}$ can be derived by applying the composition rule of
Eq.~(\ref{compRULE}) and noting that $\sqrt{\rho}$ has $G/2$. Thus, we have%
\begin{equation}
\exp\left(  i\Omega G_{\mathrm{tot}}\right)  =\exp\left(  \frac{i\Omega G_{1}%
}{2}\right)  \exp\left(  i\Omega G_{2}\right)  \exp\left(  \frac{i\Omega
G_{1}}{2}\right)  . \label{GtotDECO}%
\end{equation}

Using the expression of the partition function $Z_{\rho}$\ in
Eq.~(\ref{Gexpression}), the relation between the CM $V$\ and the Gibbs matrix
in Eq.~(\ref{e.GtoV}) into $\mathcal{F}(\hat{\rho}_{1},\hat{\rho}%
_{2})=Z_{\sqrt{\rho_{\mathrm{tot}}}}/\sqrt{Z_{\rho_{1}}Z_{\rho_{2}}}$, we may
write%
\begin{align}
\mathcal{F}(\hat{\rho}_{1},\hat{\rho}_{2})^{-4}  &  =\frac{\det\left(
e^{i\Omega\,G_{\mathrm{tot}}/4}-e^{-i\Omega\,G_{\mathrm{tot}}/4}\right)
\det\left(  e^{i\Omega\,G_{\mathrm{tot}}/4}-e^{-i\Omega\,G_{\mathrm{tot}}%
/4}\right)  }{\det\left(  e^{i\Omega\,G_{1}/2}-e^{-i\Omega\,G_{2}/2}\right)
\det\left(  e^{i\Omega\,G_{2}/2}-e^{-i\Omega\,G_{2}/2}\right)  }\nonumber\\
&  =\frac{\det\left(  e^{i\Omega\,G_{\mathrm{tot}}/2}-\openone\right)
\det\left(  e^{i\Omega\,G_{\mathrm{tot}}/2}-\openone\right)  }{\det\left(
e^{i\Omega\,G_{1}}-\openone\right)  \det\left(  e^{i\Omega\,G_{2}%
}-\openone\right)  }\nonumber\\
&  =\frac{\det\left(  e^{i\Omega\,G_{\mathrm{tot}}}-\openone\right)  }%
{\det\left(  e^{i\Omega\,G_{1}}-\openone\right)  \det\left(  e^{i\Omega
\,G_{2}}-\openone\right)  }\frac{\det\left(  e^{i\Omega\,G_{\mathrm{tot}}%
/2}-\openone\right)  }{\det\left(  e^{i\Omega\,G_{\mathrm{tot}}/2}%
+\openone\right)  }\nonumber\\
&  =\frac{\det\left(  e^{i\Omega\,G_{2}}-e^{-i\Omega\,G_{1}}\right)  }%
{\det\left(  \openone-e^{-i\Omega\,G_{1}}\right)  \det\left(  e^{i\Omega
\,G_{2}}-\openone\right)  }\frac{\det\left(  e^{i\Omega\,G_{\mathrm{tot}}%
/2}-\openone\right)  }{\det\left(  e^{i\Omega\,G_{\mathrm{tot}}/2}%
+\openone\right)  }\nonumber\\
&  =\left[  \Gamma(G_{1},G_{2})~F_{\mathrm{tot}}\right]  ^{-4}
\label{fidGAMMA}%
\end{align}
where%
\begin{align}
\Gamma(G_{1},G_{2})  &  :=\sqrt[4]{\frac{\det\left(  \openone-e^{-i\Omega
\,G_{1}}\right)  \det\left(  e^{i\Omega\,G_{2}}-\openone\right)  }{\det\left(
e^{i\Omega\,G_{2}}-e^{-i\Omega\,G_{1}}\right)  \det{i\Omega}}},\nonumber\\
F_{\mathrm{tot}}  &  :=\sqrt[4]{\frac{\det\left(  e^{i\Omega\,G_{\mathrm{tot}%
}/2}+\openone\right)  }{\det\left(  e^{i\Omega\,G_{\mathrm{tot}}%
/2}-\openone\right)  }\det i\Omega}~. \label{FtotDERI}%
\end{align}

Now it is easy to check that
\begin{equation}
\Gamma(G_{1},G_{2})=\frac{1}{\sqrt[4]{\det[V_{1}+V_{2}]}}. \label{GammaAPP}%
\end{equation}
By contrast, the computation of $F_{\mathrm{tot}}$ is more difficult. Using
Eq.~(\ref{e.GtoW}) we may write $F_{\mathrm{tot}}$ in terms of
$W_{\mathrm{tot}}$\ as follows%
\begin{equation}
F_{\mathrm{tot}}=\det\left[  \left(  \sqrt{\openone-W_{\mathrm{tot}}^{-2}%
}+\openone\right)  W_{\mathrm{tot}}i\Omega\right]  ^{1/4}, \label{FtotAPP}%
\end{equation}
or, equivalently, in terms of $V_{\mathrm{tot}}$ as follows%
\begin{equation}
F_{\mathrm{tot}}=\det\left[  2\left(  \sqrt{\openone+\frac{(V_{\mathrm{tot}%
}\Omega)^{-2}}{4}}+\openone\right)  V_{\mathrm{tot}}\right]  ^{1/4}.
\label{e.FtotV}%
\end{equation}

Let us compute $W_{\mathrm{tot}}$ as a function of $W_{1}$ and $W_{2}$. For
this we iterate the composition rule in Eq.~(\ref{Wsec}) and we use the
following relations for the $W$-matrix of the square-root state%
\begin{align}
W_{\mathrm{sq}}  &  =\left(  \sqrt{\openone-W^{-2}}+\openone\right)
W,\label{e.wsq}\\
W  &  =\frac{1}{2}\left(  W_{\mathrm{sq}}+W_{\mathrm{sq}}^{-1}\right)  .
\label{e.w2}%
\end{align}

Let us start by applying Eq.~(\ref{Wsec}) twice. We have
\begin{align*}
W^{\prime\prime}  &  =\openone+(W_{2}-\openone)(W_{1\mathrm{sq}}+W_{2}%
)^{-1}(W_{1\mathrm{sq}}-\openone),\\
W_{\mathrm{tot}}  &  =\openone+(W_{1\mathrm{sq}}-\openone)(W_{1\mathrm{sq}%
}+W^{\prime\prime})^{-1}(W^{\prime\prime}-\openone)\\
&  =\openone+(W_{1\mathrm{sq}}-\openone)(W_{1\mathrm{sq}}+W^{\prime\prime
})^{-1}\times\\
&  \hspace{1cm}(W_{2}-\openone)(W_{1\mathrm{sq}}+W_{2})^{-1}(W_{1\mathrm{sq}%
}-\openone).
\end{align*}
Now the next step is to apply the Woodbury identity and $(A^{-1}+B^{-1}%
)^{-1}=A(A+B)^{-1}B$ multiple times, so that we have
\begin{gather*}
\lbrack W_{1\mathrm{sq}}+\openone+(W_{2}-\openone)(W_{1\mathrm{sq}}%
+W_{2})^{-1}(W_{1\mathrm{sq}}-\openone)]^{-1}=\\
(W_{1\mathrm{sq}}-\openone)^{-1}\left[  (W_{1\mathrm{sq}}+W_{2})^{-1}%
+\frac{\openone}{W_{2}-\openone}{\frac{W_{1\mathrm{sq}}+\openone}%
{W_{1\mathrm{sq}}-\openone}}\right]  ^{-1}(W_{2}-\openone)^{-1},
\end{gather*}
and we may write
\begin{align*}
W_{\mathrm{tot}}  &  =\openone+\left[  (W_{1\mathrm{sq}}+W_{2})^{-1}%
+\frac{\openone}{W_{2}-\openone}{\frac{W_{1\mathrm{sq}}+\openone}%
{W_{1\mathrm{sq}}-\openone}}\right]  ^{-1}\times\\
&  \hspace{1cm}(W_{1\mathrm{sq}}+W_{2})^{-1}(W_{1\mathrm{sq}}-\openone)\\
&  =W_{1\mathrm{sq}}-(W_{1\mathrm{sq}}+W_{2})X^{-1}(W_{1\mathrm{sq}%
}-\openone)~,
\end{align*}
where
\begin{align*}
X  &  =W_{1\mathrm{sq}}+W_{2}+\frac{W_{1\mathrm{sq}}-\openone}{W_{1\mathrm{sq}%
}+\openone}(W_2-\openone)\\&=\frac{\openone}{W_{1\mathrm{sq}}+\openone}{(}%
\openone+W_{1\mathrm{sq}}^{2}+2W_{1\mathrm{sq}}W_{2})\\
&  =\frac{W_{1\mathrm{sq}}}{W_{1\mathrm{sq}}+\openone}(W_{1\mathrm{sq}}%
^{-1}+W_{1\mathrm{sq}}+2W_{2})=\frac{2W_{1\mathrm{sq}}}{W_{1\mathrm{sq}%
}+\openone}(W_{1}+W_{2}),
\end{align*}
and we have used Eq.~\eqref{e.w2}. Therefore
\[
W_{\mathrm{tot}}=W_{1\mathrm{sq}}-\frac{1}{2}(W_{1\mathrm{sq}}+W_{2}%
)(W_{1}+W_{2})^{-1}(W_{1\mathrm{sq}}-W_{1\mathrm{sq}}^{-1}).
\]
Because $W_{1\mathrm{sq}}+W_{2}=W_{1\mathrm{sq}}+W_{2}+W_{1}-W_{1}$ and
$\frac{1}{2}(W_{1\mathrm{sq}}-W_{1\mathrm{sq}}^{-1})=W_{1\mathrm{sq}}-W_{1}$,
we may write
\[
W_{\mathrm{tot}}=W_{1}-(W_{1\mathrm{sq}}-W_{1})(W_{1}+W_{2})^{-1}%
(W_{1\mathrm{sq}}-W_{1}).
\]

This is already a simple expression, but it can be further simplified. Let us
write its inverse%
\[
W_{\mathrm{tot}}^{-1}=\frac{\openone}{W_{1\mathrm{sq}}-W_{1}}\left(
\frac{W_{1}}{(W_{1\mathrm{sq}}-W_{1})^{2}}-\frac{\openone}{W_{1}+W_{2}%
}\right)  ^{-1}\frac{\openone}{W_{1\mathrm{sq}}-W_{1}}.
\]
Using Eq.~\eqref{e.wsq} we may write%
\[
\frac{(W_{1\mathrm{sq}}-W_{1})^{2}}{W_{1}}=W_{1}-W_{1}^{-1},
\]
which, replaced in the previous expression of $W_{\mathrm{tot}}^{-1}$, leads
to%
\begin{align}
W_{\mathrm{tot}}^{-1}  &  =\frac{W_{1\mathrm{sq}}-W_{1}}{W_{1}}\left(
W_{1}-W_{1}^{-1}-W_{1}-W_{2}\right)  ^{-1}\nonumber\\
&  \times(W_{1}+W_{2})\frac{\openone}{W_{1\mathrm{sq}}-W_{1}}\nonumber\\
&  =-(W_{1\mathrm{sq}}-W_{1})(1+W_{2}W_{1})^{-1}(W_{1}+W_{2})\frac
{\openone}{W_{1\mathrm{sq}}-W_{1}}\nonumber\\
&  =(W_{1\mathrm{sq}}-W_{1})W_{\mathrm{aux}}^{-1}\frac{\openone}%
{W_{1\mathrm{sq}}-W_{1}}, \label{e.wtotaux}
\end{align}
where
\begin{equation}
W_{\mathrm{aux}}=-\frac{\openone}{W_{1}+W_{2}}{(}\openone+W_{2}\,W_{1}).
\label{e.wtotauxNEXT}
\end{equation}

Because in Eq.~(\ref{FtotAPP}) there is a determinant of matrix function, such
expression is invariant under $MWM^{-1}$ transformations (with non-singular
$M$).
Therefore, we can use
$W_{\mathrm{aux}}$\ in the place of $W_{\mathrm{tot}}$ in Eq.~(\ref{FtotAPP}).
In other words, we may write
\begin{align}
F_{\mathrm{tot}}  &  =\det\left[  \left(  \sqrt{\openone-W_{\mathrm{aux}}%
^{-2}}+\openone\right)  W_{\mathrm{aux}}i\Omega\right]  ^{1/4}
\label{e.FtotWa}\\
&  =\det\left[  2\left(  \sqrt{\openone+\frac{(V_{\mathrm{aux}}\Omega)^{-2}%
}{4}}+\openone\right)  V_{\mathrm{aux}}\right]  ^{1/4}, \label{e.FtotVa}%
\end{align}
where we have used $W_{\mathrm{aux}}=-2V_{\mathrm{aux}}i\Omega$. Combining
Eqs.~(\ref{fidGAMMA}), (\ref{GammaAPP}) and~(\ref{e.FtotVa}), we obtain
Eq.~(\ref{genFIDmain}), (\ref{Ftot1}) and~(\ref{Ftot2}).

\subsection{Comment for pure states}
The most important result of the previous sections is the similarity
transformation which relates $W_{\mathrm{tot}}$ and $W_{\mathrm{aux}}$:
\begin{align}
W_{\mathrm{tot}}
 =(W_{1\mathrm{sq}}-W_{1})W_{\mathrm{aux}}\frac{\openone}%
{W_{1\mathrm{sq}}-W_{1}}. \label{e.wtotaux2}%
\end{align}
However, when $\rho_1$ is pure $W_{1\mathrm{sq}}=W_{1}$ so the above
transformation is singular. The purpose of this section is to show that
the final result \eqref{e.FtotVa} is consistent even when the matrix
$W_{1\mathrm{sq}}-W_{1}$ is singular.

To simplify the notation we assume that $\rho_1$ is a pure state, so the
symplectic eigenvalues $v^1_i$ are equal $v^1_i=1/2, \forall i$, although
the following argument can be easily generalized to the case in which only
few eigenvalues are equal to $1/2$. Because Eq.\eqref{FtotAPP} is basis
independent, we perform the calculation in the basis where $W_1$ is diagonal and
we write
\begin{align}
  W_1 &= \lim_{\epsilon\to 1}W_1(\epsilon),
  &
  W_1(\epsilon) &= \epsilon D_1,
  & D_1&=\;\openone\oplus(-\openone)~.
  \label{e.w1e}
\end{align}
Since Eq.\eqref{FtotAPP} depends only on the eigenvalues of $W_{\mathrm{tot}}$
and the eigenvalues are smooth under perturbations we can write
\begin{equation}
  F_{\mathrm{tot}}=\lim_{\epsilon\to1}
  \det\left[  \left(  \sqrt{\openone-W_{\mathrm{tot}(\epsilon)}^{-2}%
}+\openone\right)  W_{\mathrm{tot}(\epsilon)}i\Omega\right]  ^{1/4},
\label{FtotAPPe}%
\end{equation}
where $
W_{\mathrm{tot}(\epsilon)}$ refers to $W_{\mathrm{tot}}$ with $W_1$
substituted by $W_1(\epsilon)$.
For any $\epsilon<1$, it is $W_{1\mathrm{sq}}-W_{1}=\sqrt{1-\epsilon^{-2}} D_1$
so the similarity transform \eqref{e.wtotaux2} is well defined and
\eqref{FtotAPPe} can be replaced by \eqref{e.FtotWa}.
Although the matrix $W_{1\mathrm{sq}}-W_{1}$ is singular for $\epsilon\to1$ its
dependence cancels out, while $W_{\mathrm{aux}}$ is well-defined even in
the limit $\epsilon\to1$.

This is confirmed by the fact that \eqref{FtotAPP} reproduces the known results
\cite{spedalieri_limit_2013} when $\rho_1$ is pure.
In the next section we expand this point to simplify the numerical treatment of
the singular case.

\subsection{Treatment of the singular case}
In this section we devise a strategy that helps
the numerical treatment of the singular case, i.e. when one or more
symplectic eigenvalues of $V_1$ and/or $V_2$ are equal to 1/2.
Because the eigenvalues of $W_{\mathrm{aux}}$ are invariant under the exchange
of the states $\rho_1\leftrightarrow \rho_2$, without loss of generality we assume that
$V_1$ is the state with the highest number of eigenvalues equal to $1/2$.
Let $r$ be the number of pairs of symplectic eigenvalues of $V_1$ equal to
$1/2$.
Since $V_{\mathrm{aux}}$ transforms under symplectic
transformations, without loss of generality we can perform the calculations in
the coordinate system where $V_1$ is diagonal.

Moreover, to simplify the notation, in this section we reshape the matrices so
that $\Omega=\oplus_j \begin{pmatrix}
  0&1\\-1&0
\end{pmatrix}$.
Therefore, we can can write $V_1$ and $V_2$ in the block form where
\begin{align}
  V_1&=\begin{pmatrix}
  \openone_{2r}/2 & 0 \\ 0 & D
  \end{pmatrix}~,
  &V_2&=\begin{pmatrix}
  A & C \\ C^T & B
  \end{pmatrix}~,
  &
  \Omega&=\begin{pmatrix}
  \omega & 0 \\ 0 & \tilde\omega
  \end{pmatrix}~,
\end{align}
where $\openone_{2r}, A, \omega$ are $2r\times2r$ matrices, $C$ is a
$2r\times2(n-r)$ matrix, and $D, B, \tilde\omega$ are $2(n-r)\times2(n-r)$
matrices, $D$ is diagonal with diagonal entries greater then $1/2$.
Thanks to this block structure, with a long but straightforward calculation
we find
\begin{align}
  V_{\mathrm{aux}}= \begin{pmatrix}
    \openone_{2r}/2 & \tilde{C} \\ 0 &\tilde{B}
  \end{pmatrix}~,
  \label{e.vauxblock}
\end{align}
where the matrices $\tilde{C}$ and $\tilde B$ depend on $A,B,C,D$.
Because of the block structure of Eq.~\eqref{e.vauxblock}, it is
clear that
$W_{\mathrm{aux}}$ has $r$ eigenvalues equal to 1 and $r$
eigenvalues equal to $-1$. In view of Eq.~\eqref{FtotEIGEN}, these
eigenvalues do not contribute to $F_{\mathrm{tot}}$ and can thus
be discarded. On the other hand, the eigenvalues
$w_j^{\mathrm{aux}}\neq\pm1$ can be found by diagonalizing
$\tilde{W}_{\mathrm{aux}}=-2i\tilde B\tilde\omega$. With a similar
argument, $I_{2k}=2r+\Tr[\tilde{W}_{\mathrm{aux}}^{2k}]$.

\subsection{Alternative Formula\label{altformulaAPP}}

Note that in the proof of Sec.~\ref{ProofAPP}\ we can exploit the fact that
$\det[f(V)]=\det[f(UVU^{-1})]$ for some invertible matrix $U$. By using
Eq.~(\ref{GtotDECO}) into Eq.~(\ref{FtotDERI}), we get
\[
F_{\mathrm{tot}}^{4}=\frac{\det\left[  \sqrt{e^{i\Omega\,G_{1}/2}%
e^{i\Omega\,G_{2}}e^{i\Omega\,G_{1}/2}}+\openone\right]  }{\det\left[
\sqrt{e^{i\Omega\,G_{1}/2}e^{i\Omega\,G_{2}}e^{i\Omega\,G_{1}/2}%
}-\openone\right]  }\det i\Omega,
\]
and with either $U=e^{i\Omega\,G_{1}/2}$ or $U=e^{-i\Omega\,G_{1}/2}$
\begin{align}
F_{\mathrm{tot}}^{4}  &  =\frac{\det\left[  \sqrt{e^{i\Omega\,G_{2}}%
e^{i\Omega\,G_{1}}}+\openone\right]  }{\det\left[  \sqrt{e^{i\Omega\,G_{2}%
}e^{i\Omega\,G_{1}}}-\openone\right]  }\det i\Omega\nonumber\\
&  =\frac{\det\left[  \sqrt{e^{i\Omega\,G_{1}}e^{i\Omega\,G_{2}}%
}+\openone\right]  }{\det\left[  \sqrt{e^{i\Omega\,G_{1}}e^{i\Omega\,G_{2}}%
}-\openone\right]  }\det i\Omega.
\end{align}
Finally, after simple algebra, we may write
\begin{align}
F_{\mathrm{tot}}^{4}  &  =\det\left[  2\left(  \sqrt{\openone+\frac
{(V_{12}\Omega)^{-2}}{4}}+\openone\right)  V_{12}\right] \label{e.FtotV12}
\\
&  =\det\left[  2\left(  \sqrt{\openone+\frac{(V_{21}\Omega)^{-2}}{4}%
}+\openone\right)  V_{21}\right]  , \label{e.FtotV21}%
\end{align}
being
\[
V_{12}=-\frac{i\Omega}{2}+\left(  V_{1}+\frac{i\Omega}{2}\right)  (V_{1}%
+V_{2})^{-1}\left(  V_{2}+\frac{i\Omega}{2}\right)  ,
\]
and $V_{21}=V_{12}^{\dagger}$. Note that, contrary to matrix $V_{\mathrm{aux}%
}$, the new matrix $V_{12}$ is not real.
Because of the above derivation,
$W_{\rm tot}= e^{i\Omega\,G_{1}/2}W_{12}
e^{-i\Omega\,G_{1}/2}$, and
$W_{\rm tot}= e^{-i\Omega\,G_{1}/2}W_{21}
e^{i\Omega\,G_{1}/2}$ so the matrices $W_{\rm tot}$, $W_{12}$ and $W_{21}$ are
similar.

The relation between $V_{12}$ and $V_{\mathrm{aux}}$ is easy to obtain using
the $W$ matrices and applying the Woodbury identity. We find
\begin{align}
W_{12}  &  =\openone+(W_{1}-\openone)(W_{1}+W_{2})^{-1}(W_{2}-\openone)\nonumber\\
&  =(W_{1}-\openone)\left(  \frac{\openone}{W_{2}W_{1}-W_{1}-W_{2}+\openone}
+\frac{\openone}{W_{1}+W_{2}}\right)  (W_{2}-\openone)\nonumber\\
&  = (W_{2}-1)^{-1}(W_{2}W_{1}+\openone)\frac{\openone}{W_{1}+W_{2}}%
(W_{2}-\openone)\nonumber\\
&  = (W_{1}-1)\frac{\openone}{W_{1}+W_{2}}(W_{2}W_{1}+\openone)(W_{1}%
-\openone)^{-1}%
\label{e.passaggioW12aux}
\end{align}
so that $W_{12}=-UW_{\mathrm{aux}}U^{-1}$ for some invertible $U$,
as we can see by comparing Eq.~(\ref{e.passaggioW12aux}) with
Eq.~(\ref{e.wtotauxNEXT}).

\subsection{Exchanging $\rho_1$ and $\rho_2$}
The final result for the fidelity, Eq.~$\eqref{e.FtotWa}$, depends
on the matrix $W_{\mathrm{aux}}$ which is not symmetric upon
exchanging $\rho_1$ and $\rho_2$. This is due to the apparent
asymmetry in the definition of the fidelity \eqref{fid2}. However,
we show here that \eqref{e.FtotWa} is invariant under such
exchange, even though $W_{\mathrm{aux}}$ is not. Indeed, thanks to
the results of the previous section, if $F(W) =
\det\left[\left(\sqrt{\openone-W^{-2}}+\openone\right)
W\right]^{1/4}$, then $F_{\mathrm{tot}}=F(W_{\mathrm{tot}})
=F(W_{\mathrm{aux}}) =F(W_{12}) =F(W_{21})$. Because
$W_{\mathrm{aux}}$ is similar to $W_{12}$ (apart from a global
sign), which again is similar $W_{21}$, if we exchange $\rho_1$
and $\rho_2$, the resulting $W_{\mathrm{aux}}$ (with indices 1 and
2 swapped) is similar to the original one. Therefore,
\eqref{e.FtotWa} is invariant under such exchange.

\subsection{Derivation of the fidelity for displaced Gaussian states}
\label{displAPP}

Consider displaced Gaussian states, $\rho_{1}$ having Gibbs matrix $G_{1}$ and
mean value $u_{1}$, and $\rho_{2}$, having $G_{2}$ and $u_{2}$. Then
\begin{align*}
\mathcal{F}(\hat{\rho}_{1},\hat{\rho}_{2})  &  =\frac{Z_{\sqrt{\rho
_{\mathrm{tot}}}}}{\sqrt{Z_{\rho_{1}}Z_{\rho_{2}}}}=\frac{Z_{\sqrt
{\rho_{G_{\mathrm{tot}}}}}}{\sqrt{Z_{\rho_{G_{1}}}Z_{\rho_{G_{2}}}}}%
\frac{Z_{\sqrt{\rho_{\mathrm{tot}}}}}{Z_{\sqrt{\rho_{G_{\mathrm{tot}}}}}},\\
&  =\mathcal{F}(\hat{\rho}_{G_{1}},\hat{\rho}_{G_{2}})\frac{Z_{\sqrt
{\rho_{\mathrm{tot}}}}}{Z_{\sqrt{\rho_{G_{\mathrm{tot}}}}}},
\end{align*}
where $\mathcal{F}(\hat{\rho}_{G_{1}},\hat{\rho}_{G_{2}})$ is the fidelity
(already computed) between two undisplaced Gaussian states, i.e., with Gibbs
matrices $G_{1}$ and $G_{2}$ but zero mean values. Therefore, we only need to
compute $Z_{\sqrt{\rho_{\mathrm{tot}}}}/Z_{\sqrt{\rho_{G_{\mathrm{tot}}}}}$.
If we write
\[
\rho_{\mathrm{tot}}=e^{-\frac{1}{2}(Q-u_{\mathrm{tot}})^{T}\,G_{\mathrm{tot}%
}\,(Q-u_{\mathrm{tot}})+K_{\mathrm{tot}}}~
\]
then
\[
Z_{\sqrt{\rho_{\mathrm{tot}}}}=Z_{\sqrt{\rho_{G_{\mathrm{tot}}}}%
}e^{K_{\mathrm{tot}}/2}~.
\]
Moreover, from the definition one can see that
\begin{equation}
e^{K_{\mathrm{tot}}}=\frac{Z_{\rho_{\mathrm{tot}}}}{Z_{\rho_{G_{\mathrm{tot}}%
}}}=\frac{Z_{\rho_{1}\rho_{2}}}{Z_{\rho_{G_{1}}\rho_{G_{2}}}}~. \label{eKtot}%
\end{equation}
For the numerator we may write
\begin{align*}
Z_{\rho_{1}\rho_{2}}  &
=\Tr[\rho_1\rho_2]=\Tr[D(-u_1) \rho_{G_1} D(u_1)D(-u_2) \rho_{G_2} D(u_2)]\\
&  =\Tr[D(u_2-u_1) \rho_{G_1} D(u_1-u_2) \rho_{G_2}]
\end{align*}
where the phase in Eq.~\eqref{e.iddisplaced} vanishes after the twofold use.
Then calling
\[
\delta_{u}=u_{2}-u_{1},
\]
and calling $G_{12}$ the matrix such that
\[
e^{-i\Omega G_{12}}=e^{-i\Omega G_{1}}e^{-i\Omega G_{2}},
\]
one has
\begin{align*}
Z_{\rho_{1}\rho_{2}}  &
=\Tr[e^{\delta_u^T i\Omega Q} \rho_{G_1} e^{-\delta_u^T i\Omega Q}\rho_{G_2}]\\
&
=\Tr[e^{\delta_u^T i\Omega Q} e^{-\delta_u^T i\Omega e^{i\Omega G_1}Q} \rho_{G_1} \rho_{G_2}]\\
&
=\Tr[e^{\delta_u^T i\Omega Q} e^{-(e^{-i\Omega G_1}\delta_u)^T i\Omega Q} \rho_{G_{12}}]\\
&
=\Tr[e^{(\delta_u-e^{-i\Omega G_1}\delta_u)^T i\Omega Q} e^{\frac12 \delta_u^Ti\Omega e^{-i\Omega G_1}\delta_u} \rho_{G_{12}}].
\end{align*}
Now, by using Eq.~\eqref{e.complinear} we find%
\[
Z_{\rho_{1}\rho_{2}}=e^{\frac{1}{2}\delta_{u}^{T}i\Omega e^{-i\Omega G_{1}%
}\delta_{u}}e^{\frac{1}{4}(\delta_{u}-e^{-i\Omega G_{1}}\delta_{u})^{T}i\Omega
W_{12}(\delta_{u}-e^{-i\Omega G_{1}}\delta_{u})}\Tr[\rho_{G_1}\rho_{G_2}].
\]
By replacing the latter expression into Eq.~(\ref{eKtot}), we derive
\begin{align*}
e^{K_{\mathrm{tot}}}  &  =e^{\frac{1}{2}\delta_{u}^{T}i\Omega e^{-i\Omega
G_{1}}\delta_{u}}e^{\frac{1}{4}(\delta_{u}-e^{-i\Omega G_{1}}\delta_{u}%
)^{T}i\Omega W_{12}(\delta_{u}-e^{-i\Omega G_{1}}\delta_{u})}\\
&  =e^{\frac{1}{4}\delta_{u}^{T}i\Omega(e^{-i\Omega G_{1}}-e^{i\Omega G_{1}%
})\delta_{u}}e^{\frac{1}{4}\delta_{u}^{T}i\Omega(1-e^{i\Omega G_{1}}%
)W_{12}(1-e^{-i\Omega G_{1}})\delta_{u}}.
\end{align*}

The term $K_{\mathrm{tot}}$ can be simplified noting that%
\begin{gather*}
e^{-i\Omega G_{1}}-e^{i\Omega G_{1}}=\frac{W_{1}+\openone}{W_{1}%
-\openone}-\frac{W_{1}-\openone}{W_{1}+\openone}=\frac{4W_{1}}{W_{1}%
^{2}-\openone},\\
(1-e^{i\Omega G_{1}})W_{12}(1-e^{-i\Omega G_{1}})=-\frac{2}{W_{1}%
+\openone}W_{12}\frac{2}{W_{1}-\openone},
\end{gather*}
and $W_{12}=\openone+W_{1}-\openone-(W_{1}+\openone)(W_{1}+W_{2})^{-1}%
(W_{1}-\openone)$, which is a consequence of the identity
\eqref{dyson}. Therefore, we
may write
\[
K_{\mathrm{tot}}=\delta_{u}^{T}i\Omega(W_{1}+W_{2})^{-1}\delta_{u}=-\frac
{1}{2}\delta_{u}^{T}(V_{1}+V_{2})^{-1}\delta_{u},
\]
and finally
\[
\mathcal{F}(\hat{\rho}_{1},\hat{\rho}_{2})=\frac{F_{\mathrm{tot}}}{\left(
\det[V_{1}+V_{2}]\right)  ^{1/4}}e^{-\frac{1}{4}\delta_{u}^{T}(V_{1}%
+V_{2})^{-1}\delta_{u}}.
\]

\section{Proof that the solutions for the three-mode case are real}
\label{RealSolu}
As written in \eqref{e.I3}, in the three-mode case the characteristic polynomial
\eqref{e.charpol} can be written as
  $\chi = t^3+pt+q$.
The equation $\chi=0$ has real solutions if $p<0$ and $q^2/4+p^3/27<0$,
which is simple to prove.
\def\wwauxa{(w^{\rm aux}_1)}
\def\wwauxb{(w^{\rm aux}_2)}
\def\wwauxc{(w^{\rm aux}_3)}
Indeed, calling $\pm w^{\rm aux}_i$ the eigenvalues of $W^{\rm
aux}$ one finds that $I_{2n}= 2 [\wwauxa^{2n}+\wwauxb^{2n}+\wwauxc^{2n}]$. Hence
\begin{align}
  p&=-\frac13\Big[ \wwauxa^4+\wwauxb^4+\wwauxc^4
    \nonumber
  \\&\phantom{=-\frac13(}-\wwauxa^2\wwauxb^2
  \nonumber
  -\wwauxa^2\wwauxc^2 -\wwauxb^2\wwauxc^2 \Big]
  \\&=-\frac16\Big[ \left( \wwauxa^2-\wwauxb^2\right)^2
    \nonumber
  +\left( \wwauxa^2-\wwauxc^2\right)^2
  \\&\phantom{=-\frac13(}
  +\left( \wwauxb^2-\wwauxc^2\right)^2 \Big] \le0.
\end{align}
Similarly,
\begin{align}
  \frac{q^2}4+\frac{p^3}{27}=-\frac1{108}\Big[& \nonumber
     \left( \wwauxa^2-\wwauxb^2\right)^2\left( \wwauxa^2-\wwauxc^2\right)^2\\&
   \left( \wwauxb^2-\wwauxc^2\right)^2 \Big] \le0~.
\end{align}
Hence, the eigenvalues of $W^{\rm aux}$ are real.
The real solutions of $\chi=0$ are given by \eqref{e.waux3}.

\section{Derivation of the Bures metric\label{BuresAPP}}

Let us consider two infinitesimally-close Gaussian states $\hat{\rho}_{1}%
=\hat{\rho}$ and $\hat{\rho}_{2}=\hat{\rho}+d\hat{\rho}$. The first state is
parametrized by $G$ (or $V$) and $u$, while the second state is parametrized
by $G+dG$ (or $V+dV$) and $u+du$. Hence, up to the second order
\[
  \frac{\openone}{V_{1}+V_{2}}=\frac{\openone}{{2V}}{\frac{\openone}%
{\openone+\frac{dV}{2V}}}\simeq\frac{\openone}{2V}-\frac
{\openone}{2V}dV\frac{\openone}{2V}+\frac{\openone}{2V}dV\frac
{\openone}{2V}dV\frac{\openone}{2V},
\]
and%
\[
\delta_{u}^{T}(V_{1}+V_{2})^{-1}\delta_{u}\simeq du^{T}V^{-1}du/2~.
\]

In a similar way, we find
\begin{align*}
-W_{\mathrm{aux}}^{-1}  &  =\frac{1}{\openone+W^{2}+dW\,W}(2W+dW)\\
&  =\frac{2W}{1+W^{2}}-\frac{1}{1+W^{2}}dW\frac{W^{2}-1}{W^{2}+1}+\\
&  \quad+\frac{1}{1+W^{2}}dW\frac{W}{1+W^{2}}dW\frac{W^{2}-1}{1+W^{2}}~.
\end{align*}
Since the fidelity is an invariant, one can perform the calculations in the
basis in which $W$ is diagonal. Let us call $\tilde{W}$ the (diagonal) matrix
$W$ in this basis and $d\tilde{W}$ the corresponding infinitesimal variation
(non-diagonal). Then
\begin{align*}
-\left(  \tilde{W}_{\mathrm{aux}}^{-1}\right)  _{ij}=  &  \frac{2w_{i}}%
{1+w_{i}^{2}}\delta_{ij}-\frac{1}{1+w_{i}^{2}}d\tilde{W}_{ij}\frac{w_{j}%
^{2}-1}{1+w_{j}^{2}}+\\
&  +\sum_{k}\frac{1}{1+w_{i}^{2}}d\tilde{W}_{ik}d\tilde{W}_{kj}\frac{w_{k}%
}{1+w_{k}^{2}}\frac{w_{j}^{2}-1}{1+w_{j}^{2}}~.
\end{align*}

To expand the expression
\begin{align}
F_{\mathrm{inf}}  &  =\frac{F_{\mathrm{tot}}^{4}}{\det(V_{1}+V_{2}%
)}\nonumber\\
&  =\frac{\det{W_{\mathrm{aux}}}}{\det(W+dW/2)}\det\left(  \sqrt
{\openone-W_{\mathrm{aux}}^{-2}}+\openone\right)  , \label{e.FexprBures}%
\end{align}
one has to expand
\begin{equation}
\sqrt{\openone-W_{\mathrm{aux}}^{-2}}=K^{(0)}+K^{(1)}+K^{(2)}
\label{e.Fexansion}%
\end{equation}
in terms of the $0^{\text{th}}$ order, first order and second order operators
$K^{(n)}$. Taking the square of Eq.~\eqref{e.Fexansion} and calling
\[
W_{\mathrm{aux}}^{-1}=V^{(0)}+V^{(1)}+V^{(2)}%
\]
the $2^{\text{nd}}$ order expansion of $W_{\mathrm{aux}}^{-1}$, we find the
relations
\begin{align*}
K_{(0)}^{2}  &  =\openone-V_{(0)}^{2}\\
K_{(1)}K_{(0)}+K_{(0)}K_{(1)}  &  =-V_{(1)}V_{(0)}-V_{(0)}V_{(1)}\\
K_{(2)}K_{(0)}+K_{(0)}K_{(2)}  &  =-V_{(2)}V_{(0)}-V_{(0)}V_{(2)}-V_{(1)}%
^{2}-K_{(1)}^{2}~.
\end{align*}

These explicit calculation of $K^{(n)}$ is long but straightforward. Once the
operators $K$ are known, from the expansion
\[
\det(\openone+X)=e^{\Tr\log(1+X)}\simeq e^{\Tr[X]-\Tr[X^2]/2}%
\]
of the three terms in Eq.~\eqref{e.FexprBures}, we find
\[
F_{\mathrm{inf}}=\exp\left(  \frac{1}{4}\sum_{ij}\frac{d\tilde{W}_{ij}%
d\tilde{W}_{ji}}{1-w_{i}w_{j}}\right)  ,
\]
i.e.
\[
\mathcal{F}(\rho,\rho+d\rho)=\exp\left(  -\frac{1}{8}duV^{-1}du+\frac{1}%
{16}\sum_{ij}\frac{d\tilde{W}_{ij}d\tilde{W}_{ji}}{1-w_{i}w_{j}}\right)  .
\]

The Bures metric is then given by
\[
ds^{2}=\frac{1}{4}du^{T}\,V^{-1}\,du+\frac{1}{8}\sum_{ij}\frac{d\tilde{W}%
_{ij}d\tilde{W}_{ji}}{w_{i}w_{j}-1}~.
\]
The above expression can be cast into a basis-independent form by defining the
super-operator
\[
\mathcal{L}_{A}X=AXA~.
\]
Indeed
\begin{align*}
\sum_{ij}\frac{d\tilde{W}_{ij}d\tilde{W}_{ji}}{w_{i}w_{j}-1}  &  =\Tr\left[
d\tilde{W}\frac{1}{\mathcal{L}_{\tilde{W}}-1}d\tilde{W}\right] \\
&  =\Tr\left[  dW\frac{1}{\mathcal{L}_{W}-1}dW\right] \\
&  =-4\Tr\left[  dV\Omega\frac{1}{\mathcal{L}_{W}-1}(dV\Omega)\right]  .
\end{align*}
Using
\[
\mathcal{L}_{W}(dV\Omega)=-4V\Omega dV\Omega V\Omega=-4(\mathcal{L}%
_{V}\mathcal{L}_{\Omega}dV)\Omega~,
\]
we find
\begin{align*}
\sum_{ij}\frac{d\tilde{W}_{ij}d\tilde{W}_{ji}}{w_{i}w_{j}-1}  &  =4\Tr\left[
dV\Omega\frac{1}{4\mathcal{L}_{V}\mathcal{L}_{\Omega}+1}(dV)\Omega\right] \\
&  =4\Tr\left[  dV\mathcal{L}_{\Omega}\frac{1}{4\mathcal{L}_{V}\mathcal{L}%
_{\Omega}+1}dV\right] \\
&  =4\Tr\left[  dV\frac{1}{\mathcal{L}_{\Omega}}\frac{1}{4\mathcal{L}%
_{V}\mathcal{L}_{\Omega}+1}dV\right] \\
&  =4\Tr\left[  dV\frac{1}{4\mathcal{L}_{V}+\mathcal{L}_{\Omega}}dV\right]  ,
\end{align*}
where we have used $\mathcal{L}_{\Omega}^{2}=1$. Finally, we may write%
\[
ds^{2}=\frac{1}{4}du^{T}\,V^{-1}\,du+\frac{1}{2}\Tr\left[  dV\frac
{1}{4\mathcal{L}_{V}+\mathcal{L}_{\Omega}}dV\right]  .
\]

\subsection{Singular case}
In the singular case, i.e. when some of the eigenvalues of $W$ are $\pm1$, the
sum in \eqref{BuresBasis} is performed only along the elements where $w_i
w_j\neq1$. The proof of this fact  closely follows an analogous observation in the
fermionic case \cite{Fermion}.  Let $W=\sum_i w_i \ket i \bra i$ be the eigenvalue
decomposition of $W$, where $\ket i$ is the eigenvector of $W$ with eigenvalue
$w_i$ and let $c_i=w_i^{-1}\in[-1,1]$, $c_i=\tanh(g_i/2)$, where $g_i$ are the
symplectic eigenvalues of $G$.
Using this notation,
$dW = \sum_i (1-w_i^2)\frac{dg_i}2\ket i \bra i+ w_i(\ket i\bra{di}+\ket{di}\bra
i)$. Inserting the above expression in
\eqref{BuresBasis} we find
\begin{align*}
  \delta&:= \Tr\left[  dW\frac{1}{\mathcal{L}_{W}-1}dW\right]
  \\&
  = \frac14 \sum_i (1-w_i^2)dg_i^2 + \sum_{i\neq j}
  \frac{(w_i-w_j)^2}{1-w_iw_j} |\langle di|j\rangle|^2~.
\end{align*}
The first term in the above equation is well-defined also when $w_i\to\pm1$.
To prove that the second term is bounded we define $f(x,y)=(x-y)^2(1-xy)^{-1}$
and write
\begin{align}
  \delta
  = \frac14 \sum_i (1-w_i^2)dg_i^2 + \sum_{i\neq j} f(c_i, c_j)w_i w_j
  |\langle di|j\rangle|^2~.
  \label{buressum}
\end{align}
As shown in Lemma 3 of Ref.\cite{Fermion}, the function $f(x,y)$ is bounded
in $[-1,1]^2$, $f(x,y)\le4$,  and $\lim_{(x,y)\to(\pm1,\pm1)}f(x,y)=0$.
Therefore, the elements such that $w_iw_j=1$ do not contribute in the sum
\eqref{buressum}. Numerically, this corresponds to taking the pseudo-inverse
of the superoperator in \eqref{BuresBasis} or, equivalently, in manually
avoiding the sum over the elements such that $w_iw_j=1$.
Therefore, even though Eq.\eqref{buressum} has been found assuming that
$w_i\neq\pm1$, it can be analytically extended to the general case.

Notice that for pure states, where $w_i=\pm1$, the effect of the function
$f$ in \eqref{buressum} can be obtained equivalently by the function
$\tilde f(x,y)=(x-y)^2/2$. Taking this substitution in \eqref{buressum} we find
\begin{align}
  \delta_{\mathrm{pure}} &
  = \sum_{i\neq j} \tilde f(c_i, c_j)w_i w_j
  |\langle di|j\rangle|^2~=\sum_{i,j} \frac{(w_i-w_j)^2}{2w_i w_j}
  |\langle di|j\rangle|^2~
  \nonumber\\&
  =  \frac12
  \Tr\left[  \frac{\openone}{W}\,dW\,\frac{\openone}{W}\,dW\right] ~.
  \label{deltapure}
\end{align}
The above equation provides a simpler expression for the Bures metric for
a pure Gaussian state.


\begin{thebibliography}{99}
\bibitem {BuresFID}Note that there are two definitions of quantum fidelity in
the literature. One is the square-root or Bures fidelity
$\mathcal{F}$, which is given by Eq.~(\ref{fid2}). The other is
the Uhlmann-Jozsa fidelity, which
is the squared of the previous one, i.e., $\mathcal{F}_{\text{U-J}%
}=\mathcal{F}^{2}$. Bures fidelity is the direct generalization of
the classical fidelity. In fact, if the two density operators
commute, we can write $\rho=\sum_{k}p_{k}\left\vert k\right\rangle
\left\langle k\right\vert $ and $\sigma=\sum_{k}q_{k}\left\vert
k\right\rangle \left\langle k\right\vert $ for an orthonormal
basis $\{\left\vert k\right\rangle \}$. Then, we have
$F_{B}(\rho,\sigma)=\sum_{k}\sqrt{p_{k}q_{k}}$ which is the
classical fidelity $F(p_{k},q_{k})$ between the two probability
distributions $p_{k}$ and $q_{k}$.
\bibitem{WauxNOTE} In fact, as we show in the Appendix~\ref{ProofAPP}, $W_{\mathrm{aux}}$
is equal to $W_{\mathrm{tot}}$ of the total state
$\rho_{\mathrm{tot}}$ up to a similarity transformation. This
means that $W_{\mathrm{aux}}$ and  $W_{\mathrm{tot}}$ have the
same spectrum of eigenvalues. In turn, we know that the
eigenvalues of $W_{\mathrm{tot}}$ are $\pm w_{k}^{\mathrm{tot}}$
where $w_{k}^{\mathrm{tot}}=2 \nu_{k}^{\mathrm{tot}}$ and
$\{\nu_{k}^{\mathrm{tot}}\}$ is the symplectic spectrum of
$\rho_{\mathrm{tot}}$. Thus, we have that the eigenvalues of
$W_{\mathrm{aux}}$ are $\pm w_{k}^{\mathrm{aux}}$ with
$w_{k}^{\mathrm{aux}}=2 \nu_{k}^{\mathrm{tot}}\geq1$.
\bibitem {QCRbound}Contrary to the single-parameter bound the multi-parameter
version is not necessarily achievable.
\end{thebibliography}
\end{document}